\documentclass{article}
\usepackage{PRIMEarxiv}
\usepackage{amsmath,amssymb,amsfonts}
\usepackage{mathrsfs}%
\usepackage{algorithmic}
\usepackage[]{algorithm2e}
\usepackage{array}
\usepackage[caption=false,font=normalsize,labelfont=sf,textfont=sf]{subfig}
\usepackage{textcomp}
\usepackage{stfloats}
\usepackage{url}
\usepackage{verbatim}
\usepackage{graphicx}
\usepackage{cite}

\usepackage{multirow}
\usepackage{url}
\usepackage{verbatim}
\usepackage{graphicx}
\usepackage{xcolor}%
\usepackage{textcomp}%
\usepackage{manyfoot}%
\usepackage{booktabs}%
\usepackage{listings}%
\usepackage{hyperref}
\usepackage{balance}

\begin{document}

\title{A brain-inspired generative model for EEG-based cognitive state identification$^*$}

\author{Bin Hu and Zhi-Hong Guan
\thanks{The manuscript was uploaded onto arXiv on May 3, 2025. 
This work was supported in part by the National Natural Science Foundation of China under Fund 62322311 and by the Guangdong Artificial Intelligence and Digital Economy Laboratory
(Guangzhou) under Fund PZL2023ZZ0001 (\textit{Corresponding author: Bin Hu}). }
\thanks{Bin Hu is with School of Future Technology, South China University of Technology, Guangzhou 510641, and also with Pazhou Lab, Guangzhou 510006, China. Zhi-Hong Guan with School of Artificial Intelligence and Automation, Huazhong University of Science and Technology, Wuhan 430074, China. (e-mail: huu@scut.edu.cn; zhguan@mail.hust.edu.cn).}
}

\maketitle

\begin{abstract}
This article proposes a brain-inspired generative (BIG) model that merges an impulsive-attention neural network and a variational autoencoder (VAE) for identifying cognitive states based on electroencephalography (EEG) data. A hybrid learning method is presented for training the model by integrating gradient-based learning and heteroassociative memory. The BIG model is capable of achieving multi-task objectives: EEG classification, generating new EEG, and brain network interpretation, alleviating the limitations of excessive data training and high computational cost in conventional approaches. Experimental results on two public EEG datasets with different sampling rates demonstrate that the BIG model achieves a classification accuracy above 89\%, comparable with state-of-the-art methods, while reducing computational cost by nearly 11\% over the baseline EEGNet. Incorporating the generated EEG data for training, the BIG model exhibits comparative performance in a few-shot pattern. Ablation studies justify the poised brain-inspired characteristic regarding the impulsive-attention module and the hybrid learning method. Thanks to the performance advantages with interpretable outputs, this BIG model has application potential for building digital twins of the brain. 
\end{abstract}

\section{Introduction}
\label{introduction}
Digital twins of the brain speak to the quest for generative artificial intelligence (AI), since generative models are supposed to offer interpretable and mechanistic signatures of empirical data \cite{ref-digtbrain,ref-bscale,ref-tpamidiffu}. Most existing AI models suffer from the limitations of excessive data training and high computational cost, which restrict the applications of AI-aided clinical medicine \cite{ref-shireason,ref-gmai,ref-wanghealth}. This raises one fundamental cognitive question of digital brain models, that is, how to build digital brain models to enhance cognitive states, in addition to being discriminative. 

Consider for instance the psychiatric disorders like major depressive disorder, anxiety disorder, or epilepsy. Utilizing emotion recognition or anomaly detection, AI models may provide reliable approaches to identifying these disorders. Most AI methods are data-driven, involving two steps: feature extraction and learner training \cite{ref-mlreview,ref-spikereview}. By extracting features from speech, image or EEG, data-driven models offer high performances \cite{ref-eeg2speech,ref-modma,ref-eeg2emotion}. Raw data are fed into end-to-end machine learners to produce direct diagnostic results, lacking neuronal signature and network interpretability \cite{ref-spikereview,ref-linkjzhang}. Biomedical depressive models reveal physiological abnormalities based on neural microregulation levels in human subject reports and simulated experiments in animal models \cite{ref-biodisorder}. Most biomedical models cannot be directly applied to humans for distinguishing differences in disordered symptoms and subject heterogeneity. Depressive patients suffer complicated and altered dynamics impacting spatial and temporal features of the brain activity in specific regions like anterior cingulate cortex and posterior medial frontal cortex (pMFC) \cite{ref-cortical1}. Neuroscience findings demonstrate that in human brains, pathways between pMFC and anterior midcingulate cortex (aMCC) are involved in post-error attentional adaptation, wherein links may alter emotional processing, causing psychiatric disorders \cite{ref-brainmdd,ref-cortical1,ref-cortical2}. It is thus demanded to develop neuromorphic computational models by incorporating brain-inspired characteristics of neuroplasticity and network pathways. 

\begin{figure*}
	\centering
	\includegraphics[width=0.8\textwidth]{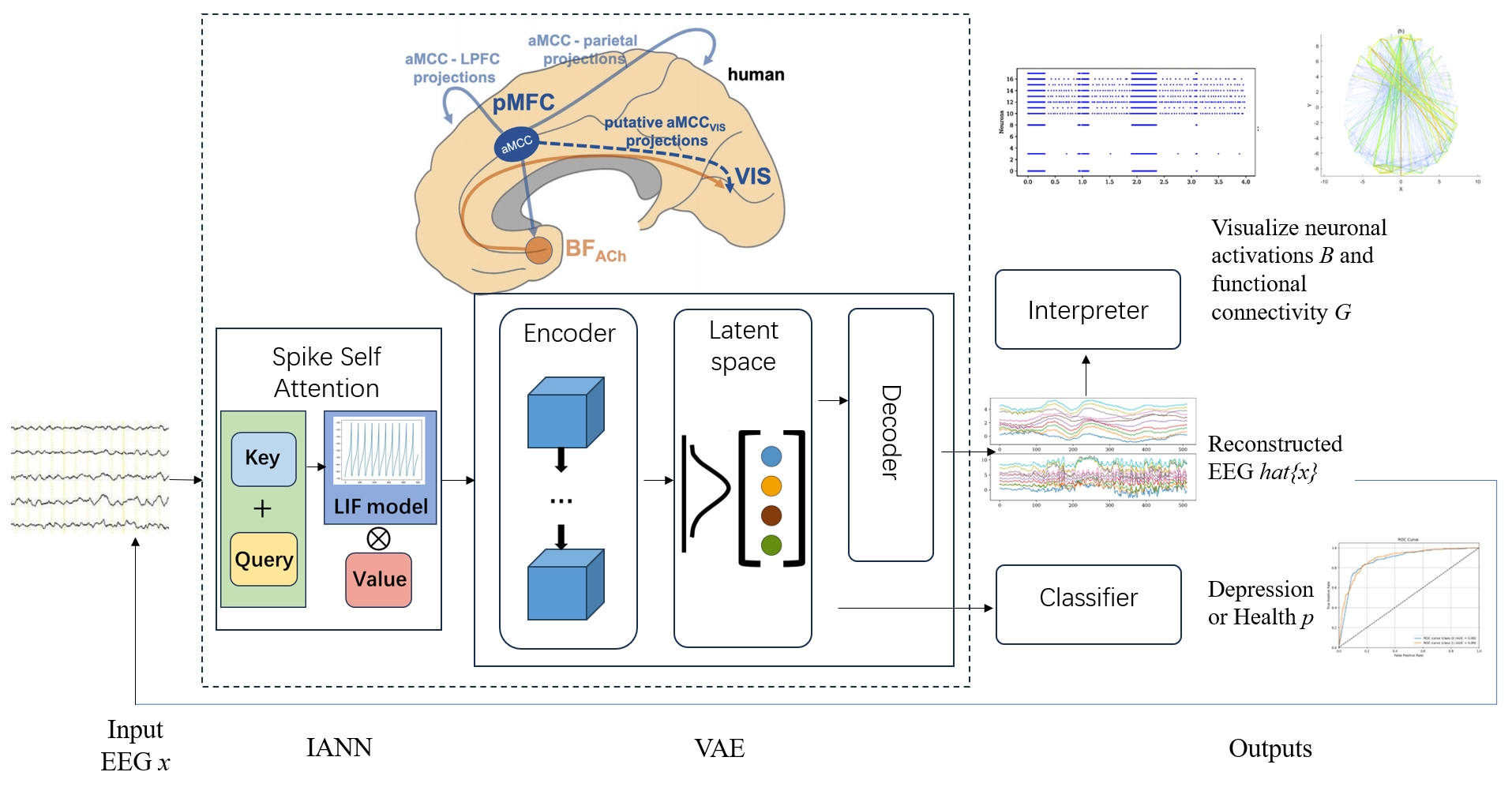}
	\caption{{The flowchart of the brain-inspired generative (BIG) model. The model integrates an IANN and a VAE by mimicking the emotional processing pathways between brain regions like pMFC and aMCC. Projections between pMFC and aMCC are involved in post-error attentional adaptation, and links therein may alter emotional processing, causing psychiatric disorders \cite{ref-cortical1,ref-cortical2}. The brain pathway at top is from \cite{ref-cortical2}}. }\label{fig1}
\end{figure*}

This article proposes a brain-inspired generative (BIG) model trained by a hybrid learning method, which can server as a computational framework for identifying cognitive states with or without depression (Fig. \ref{fig1}). An impulsive-attention neural network (IANN) and a variational autoencoder (VAE), as well as a generative brain network are developed, featuring neuroplasticity and network interpretability. The BIG model is brain-inspired in two aspects: on one hand, it integrates an IANN and a VAE by mimicking the emotional processing pathways between pMFC and aMCC \cite{ref-spikeatt,ref-memory,ref-cortical2}; on the other hand, it works as a generative brain network for  interpreting anomaly \cite{ref-brainmdd,ref-brainnetanomaly}. The hybrid learning method incorporates gradient-based global learning as a common neuromodulator, with a local associative memory to complement a personalized label \cite{ref-glocal,ref-ourmemory}. To maximize the evidence lower bound (ELBO) objective for training, the hybrid learning tends to perform multi-timescale post-error attentional adaptation as in the brain \cite{ref-cortical2,ref-fewsnn}. The BIG model offers advantages of being easily operable with low-cost in computation, which could be a computational model of digital brains through identifying cognitive states from EEG signals. 

Generative models not only can alleviate the above-mentioned two limitations of conventional AI but also can underwrite more in learnability and interpretability \cite{ref-tpamidiffu,ref-ganbrainima,ref-eeg2vec,ref-graphvae,ref-vaeimage}. In this article, we build a brain-inspired generative model with EEG inputs, enabling depression-health classification, new data generation and cognitive state interpretation. The main contributions of this article are summarized below: 
\begin{itemize}
\item Inspired by emotional pathways between pMFC and aMCC, this study combines impulsive-attention encoding with a VAE network to construct a cognitive model, which is EEG-computable and -generative, and is capable of interpreting cognitive states from the generated brain network. The proposed BIG model outperforms state-of-the-art (SOTA) approaches in two public EEG datasets. 

\item Differing from conventional VAE networks, the BIG model adopts a brain-inspired computational framework by integrating impulsive neurons and attentions for spatiotemporal encoding and utilizing a hybrid learning mechanism for training. Through maximizing the ELBO objective, this brain-inspired computation leads to improved stability in hybrid signal processing, analogous to the post-error attentional adaptation between brain regions. The BIG model automatically generates new EEG data to complement training samples, thereby alleviating the limitations of excessive data training and high computational cost. 
\end{itemize}

\section{Related Work}
\label{related_work}
This section first reviews machine learning approaches to brain anomaly detection and spiking (or impulsive) neural networks for brain-inspired computing, and then reviews generative models on neurophysiological data.

\subsection{Machine learning approaches to brain anomaly detection}\label{subsec2.1}

Previous studies have demonstrated significant progresses of machine learning in brain anomaly detection, particularly for identifying depression, epilepsy, or fatigue states \cite{ref-mlreview}. For example, a compact convolutional neural network (CNN) named EEGNet was developed for EEG-based brain-computer interfaces \cite{ref-eegnet}. Combining CNN and long short-term memory (LSTM), a hemisphere asymmetry network was established for EEG depression recognition \cite{ref-hemas}. Different types of deep ConvNets were designed for decoding imagined or executed tasks from raw EEG data \cite{ref-deepconv}. A deep CNN was studied for dealing with the EEG data from depressive and healthy subjects \cite{ref-eegcnnmdd}. A decision tree-based classifier was developed for analyzing the Patient Health Questionnaire-9 (PHQ-9) data \cite{ref-treemdd}. The support vector machine (SVM), naive Bayes and CNN were utilized for epilepsy detection \cite{ref-learnepi}. Using CNN and bi-directional LSTM, a spatiotemporal neural network was built for EEG-based emotion recognition \cite{ref-bsnn}. An epileptic seizure detection method was presented based on hybrid learning with genetic algorithm and particle swarm optimization \cite{ref-hybridepi}. Two graph attention networks were established for inferencing the brain fatigue states of pilots with EEG-based indicators \cite{ref-atteeg}. Notably, most of these methods have drawbacks in coping with data heterogeneity and interpretability. It is therefore important to build computational models offering both preciseness and interpretability, especially in application to identifying brain disorders \cite{ref-digtbrain,ref-mlreview}.

\subsection{Impulsive neural networks for brain-inspired computing}\label{subsec2.2}

Impulsive neural networks (INNs), also termed as spiking neural networks (SNNs), have emerged as one significant platform for brain-inspired computing, promising not only in high performance as AI but also in interpretable outcomes for the brain \cite{ref-spikereview,ref-networkurths}. Differing from CNNs, INNs are composed of spiking neurons and learnable synapses, carrying on hybrid dynamics of both continuous and discrete signals \cite{ref-ourmemory}. To reconstruct dynamic scenes from spikes, high-speed image reconstruction models were established using the short-term plasticity mechanism of the brain \cite{ref-huangsnn}. Both short-term learning and long-term evolution were combined to optimize few-shot learning of SNNs \cite{ref-fewsnn}. The Loihi chip was designed by simulating the impulsive learning rule of brain neural networks with multi-core SNNs \cite{ref-loihi}. SNNs were used to realize low-consumption learning with synaptic plasticity mechanisms like spike timing-dependent plasticity \cite{ref-neurosnn}. A surrogate gradient-based SNN combined with a spiking ConvLSTM unit was constructed for seizure detection \cite{ref-spikeeg}. A hybrid global-local learning model was developed, which combines a meta-learning paradigm and a differentiable spiking model with neuronal dynamics and synaptic plasticity \cite{ref-glocal}. For joint image classification and segmentation, a hybrid learning network was developed by integrating convolutional and spiking neurons and utilizing a multi-loss function for training \cite{ref-ourtai}. It is reported that SNNs have promising potentials in cognitive interpretability, due to the brain-inspired impulsive computing architecture \cite{ref-ourmemory,ref-huangsnn,ref-fewspike}. Nonetheless, there remains elusive as how to reconfigure the information processing and decision-making mechanisms of the brain. 

\subsection{Generative models for identifying neurophysiological data}\label{subsec2.3}

Prominent generative models include generative adversarial nets (GANs), variational autoencoders (VAEs), and other diffusion models \cite{ref-tpamidiffu}. GAN consists of two models trained simultaneously: a generative one and a discriminative one, and is faster at producing images than each individual one. To deal with brain image signals, for example, a pyramid of attention-based GAN and CNN was adopted to perform classification in Alzheimer's disease \cite{ref-ganbrainima}. While latent states of diffusion models have the same dimensions as the original data, VAE performs better with reduced dimensions. VAE improves also upon the conventional autoencoder by ensuring the latent space be sufficiently regularized to achieve robust training. A deep causal variational autoencoder was proposed for estimating the causal relationship between brain regions from brain disorder images \cite{ref-vaeimage}. To predict emotional states, the EEG2Vec, a conditional variational autoencoder-based framework, was developed to learn generative-discriminative representations from EEG data \cite{ref-eeg2vec}. To show EEG-based graph embedding, a graph variational autoencoder was proposed to extract nodal features from brain functional connections \cite{ref-graphvae}. A VAE and attention-based model was proposed to estimate the cognitive workload with EEG \cite{ref-cogvae}. For neurophysiological data, however, these generative models lack network interpretability and neglect the hybrid neural learning mechanism of the brain \cite{ref-networkurths,ref-glocal}.

\section{The Proposed Model}
\label{proposed_method}
This section establishes the framework and illustrates the underlying mathematics of the brain-inspired generative (BIG) model (Fig. \ref{fig1}). The model starts with an impulsive-attention neural network (IANN), used to encode EEG signals into an impulse sequence space, and then uses a VAE to further encode the impulse signals into a latent state space. In the training process, a hybrid learning mechanism is established to ensure learnability and interpretability. On the output layer, the latent state space is used to build: i) a \emph{classifier} for detecting cognitive states from an arbitrary EEG signal; ii) a \emph{generator} for simulating and reconstructing specific EEGs for feedback into the training dataset; iii) an \emph{interpreter} for mapping the latent cognitive states into brain network metrics. 

\subsection{IANN}\label{subsec3.1}

In order to cope with the complicated transmission of synchronized signals, an impulsive neural network (INN) and an attention module are merged to map EEG signals into impulse sequences. The IANN works as a spatiotemporal filter for unwrapping the neuronal discharge source signals and the collected EEG signals. IANN consists of two components: a rate-based impulse encoder and an attention processor. The EEG inputs are filtered and divided into different frequency bands, with signals from all channels in each band encoded into impulses and then fed into an attention module. 
\smallskip

\begin{figure*}[htbp]
	\centering
	\begin{minipage}{0.46\textwidth}
		\centering
		\includegraphics[width=\textwidth,height=0.165\textheight]{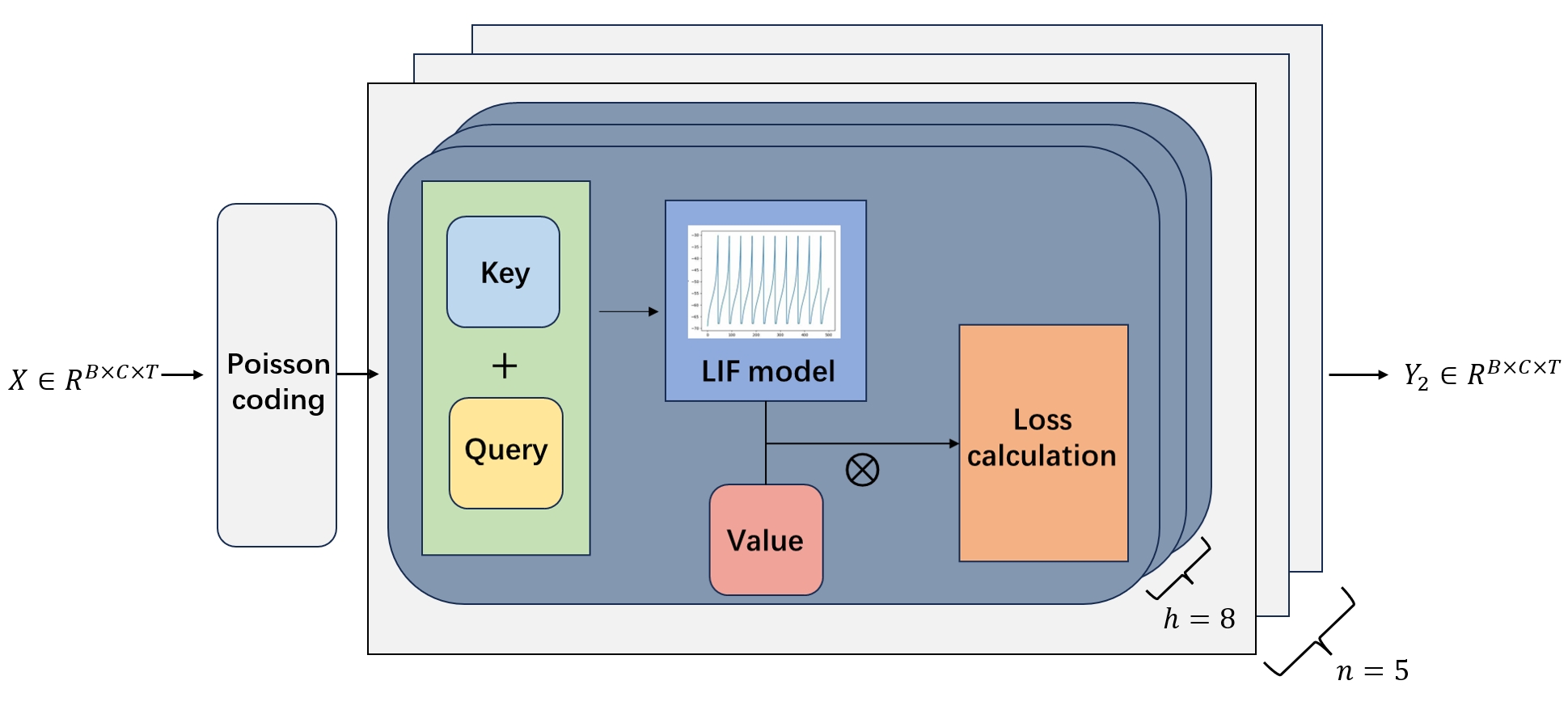} 
		\textbf{(a)} 
	\end{minipage}
	\hspace{0.2in}
	\begin{minipage}{0.46\textwidth}
		\centering
		\includegraphics[width=\textwidth,height=0.165\textheight]{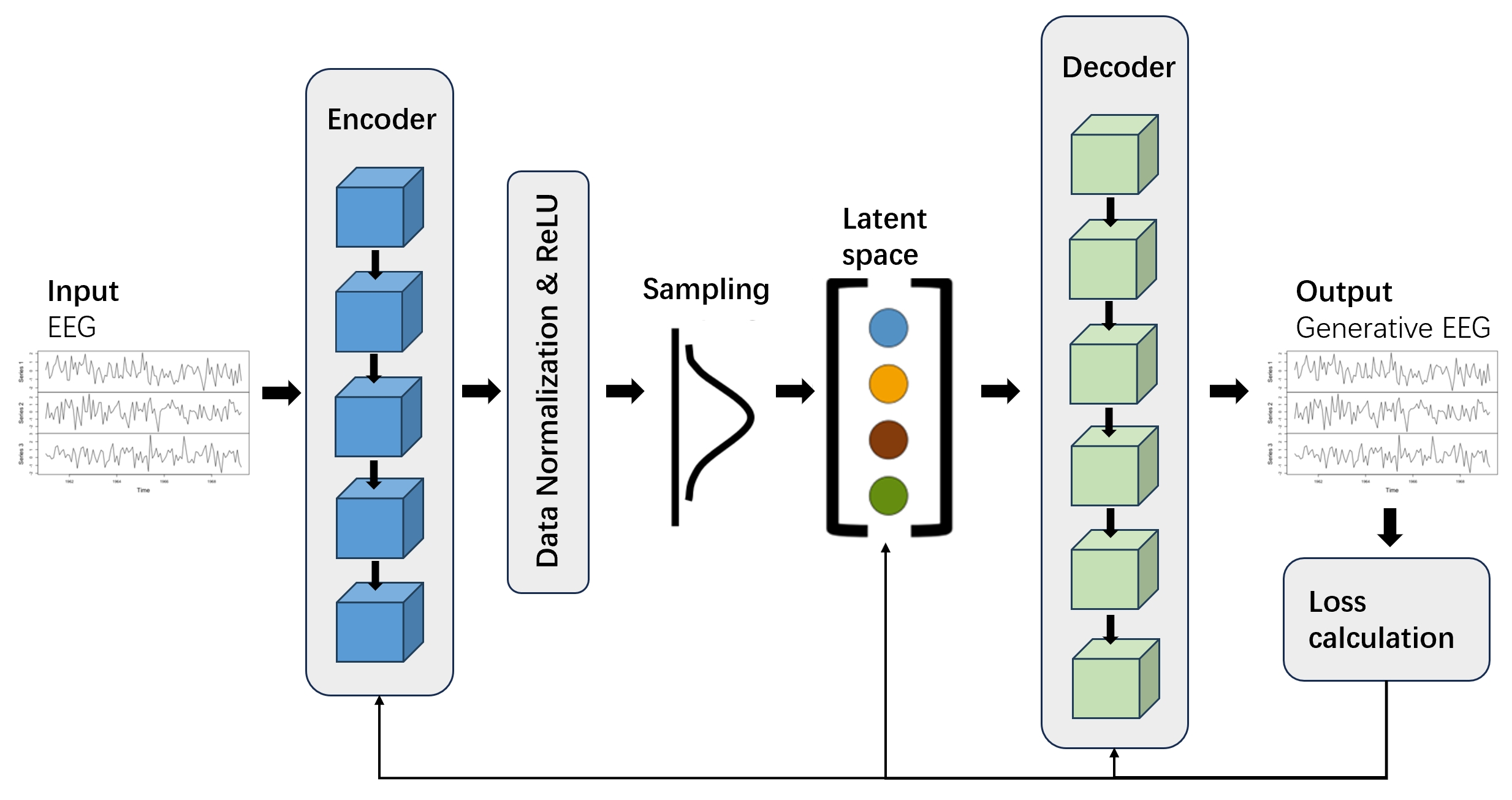} 
		\textbf{(b)} 
	\end{minipage}
	\caption{Illustrations of the impulsive-attention neural network (a) and the VAE. Panel (a) presents the IANN module using a Poisson coding, after which an LIF neuron is embedded to make an impulsive multi-head attention mechanism (8 heads), where the network has 5 hidden layers. Panel (b) shows the VAE fed with reconstructed EEG signals. Using a small amount of original EEG inputs, a latent feature space is generated after passing through a 5-layer encoder, and then a reconstructed EEG signal is generated through a 6-layer decoder.}
	\label{fig2}
\end{figure*}

A Poisson rate code is adopted for converting EEG to impulses. Define the probability of finding $\theta$ impulses in a small interval $\triangle t$ as
\begin{equation}\label{eq1}
{\rm Pro}(\theta,\triangle t)=\frac{\lambda^{\theta}}{\theta !}e^{-\lambda},
\end{equation}
where $\lambda>0$ is the mean value of impulses. Through this rate coding, the resultant impulse sequences are fed into the network. The incoming spikes, modulated by synaptic weights, cause an increase in membrane potential of neurons. Each neuron can be described by the following integrate-and-fire (LIF) model \cite{ref-huangsnn,ref-ourtai}:
\begin{equation}\label{eq2}
\tau\dfrac{\mathrm{d}u_i}{\mathrm{d}t}=-\big(u_i-u_{\rm rest}\big)+\sum_{j=1}^{N}w_{ij} \theta_j,
\end{equation}
where $u_i$ is the membrane potential, $w_{ij}$ is the synaptic weight, $\theta_j$ is the impulse count of neuron $j$, ${\tau}>0$ is a time constant, and $i=1,2,\cdots,N$. If $u_i$ exceeds threshold ${u_{\rm th}}$, neuron $i$ fires and then $u_i$ is reset to the resting potential ${u_{\rm rest}}$.
\smallskip

To cope with the heterogeneity of the EEG signals, IANN takes an attention module to enhance the spatial feature extraction. IANN utilizes impulse signals to calculate the QKV matrices in the attention module \cite{ref-spiketran}. Let $\{Q_S, K_S,V_S\}$ be the spike tensor type of the $\{Q, K,V\}$. The output of the IANN can be written as 
\begin{align}\label{eq3}
IANN(Q, K, V ) & =  g(Q_S, K_S) \otimes V_S \nonumber\\
& =  SN (SUM_c (Q_S \otimes K_S)) \otimes V_S,
\end{align}
where $\otimes$ is the Hadamard product, $g(\cdot)$ is the attention mapping, and $SUM_c(\cdot)$ is the sum of each column. The outputs of $g(\cdot)$ and $SUM_c(\cdot)$ are both row vectors, and $w_{ij}^{\rm att}$ represents the attention-weights.
\smallskip

Here, IANN contains five layers, where impulsive neurons and attention modules are integrated to generate a hybrid neuronal layer (Fig. \ref{fig2}a). Then, formulas of the impulsive neurons are determined by the original images and the preceding convolutions. First, for each layer $l$, the number of neurons $n_l$ is chosen in line with the convolution size. Second, the inputs of impulsive neurons vary across different layers. For an IANN layer $r$, the input of impulsive neuron $i$ is given by
\begin{equation}\label{eq4}
net_i^r(t)=\sum_{l=1}^{n_{r-1}} w_{ij}^{(r)}  c_j^{r-1}(t)=W_i^{(r)} c^{r-1}(t),
\end{equation}
where $w_{ij}^{(r)}$ is the weight of neurons $i,j$ on layer $l$, and $c_j^{r-1}(t)$ is the impulse accumulation over time $t$, given by
\begin{equation*} 
c_j^{r-1} (t)=\sum_{t_k\le t} \sum_k \theta(t-t_k),
\end{equation*}
where $t_k$ is the firing time and $\theta(\cdot)$ is the unit impulse function: $\theta(t-t_k)=1$ if $t=t_k$, otherwise $\theta(t-t_k)=0$. The choice of $t_k$ obeys the Poisson distribution (\ref{eq1}). Let $\mathbf{W}^{\rm iann}=\{w_{ij}^{r}\}$ be the weight matrix in the IANN. The optimization of $\mathbf{W}^{\rm iann}$  will be illustrated in the training part. 

\subsection{VAE}\label{subsec3.2}

A variational auto-encoder (VAE) is built after the IANN for latent state feature extraction and for EEG reconstruction. VAE is an unsupervised generative model, containing an inference encoder and a generative decoder for reconstruction. 
Unlike other VAEs, the adopted VAE network receives inputs from the feedforward IANN. Given an EEG signal $x^{(i)}$ with label $y^{(i)}$, a signal $\widetilde{x}$ obtained from IANN is used as input to the VAE. Here, the purposes of the VAE network are threefold: 
\begin{itemize}
	\item[(\emph{a})]
	To find a variational inference encoder $q_{\theta}(z|\widetilde{x})$ with impulse parameters $\theta$ that can generalize across EEG categories;
	\item[(\emph{b})] 
	To build a corresponding decoder that can utilize a variational distribution to synthesize new EEG data via minimizing the reconstruction error between the encoding and decoding processes;
	\item[(\emph{c})] 
	To obtain latent state features from the input EEG $X$ that are jointly representative in classifying and interpreting tasks with depression labels $y$.
\end{itemize} 
\smallskip

\textbf{\emph{Encoder for inferring latent space}.} 
The VAE-encoder takes one input layer and three convolutional (Conv) layers (Fig. \ref{fig2}b), and converts the IANN output $\widetilde{x}$ into a higher-dimensional latent state $z$: 
\begin{equation}\label{eq5}
E^{(1)}=F\left({\rm Conv}\left(\widetilde{x},W^{(1)},B^{(1)}\right)\right),
\end{equation}
where $W^{(1)},B^{(1)}$ are respectively the weight and bias matrices on the 1st layer, and $F(\cdot)$ is the ReLU activation function. The latent state $z$ is obtained from the approximate posterior distribution $q(z|\widetilde{x})$, obeying an independent Gaussian distribution $\mathcal{N}(\boldsymbol{\mu}, \text{diag}(\boldsymbol{\sigma}^2))$. It is estimated by 
\begin{equation}\label{eq6}
\left[\boldsymbol{\mu}, \text{diag}(\boldsymbol{\sigma}^2)\right]=F\left(E^{(2)}\ast W^{(2)}\right)+B^{(2)} E,
\end{equation}
which is the output of the VAE-encoder, denoted by $O_{e}$. Concerning the above objective (\emph{a}), the goal of this encoder is to solve the minimization problem formulated below:
\begin{align*}
&\min \; D_{KL} \Big(q(z|\widetilde{x})\left\|p(z)\right.\Big), \\
&\; \emph{s.t. }\;\; \mathbf{W}^{\rm e},\; \mathbf{B}^{\rm e},  
\end{align*}
where $D_{KL}(\cdot)$ is the Kullback-Leibler divergence between the approximate distribution and the true posterior, representing the loss of the inference encoder and the generative decoder, and $\mathbf{W}^{\rm e},\mathbf{B}^{\rm e}$ are respectively the required weight and bias matrices in the encoder \cite{ref-vaeimage,ref-graphvae}.
\smallskip

\textbf{\emph{Decoder for generating EEG data}.} 
The VAE-decoder takes the above adjacency matrix $O_e$ as input, and aims to reconstruct the original EEG signal $x$ through approximating the likelihood $p_{\theta}(x|z)$ of $x$ given the latent state $z$. Symmetrically to the VAE-encoder, three transposed convolutional (ConvT) layers are adopted to reconstruct the input by convolving the upsampled feature maps with filters. The formula of the decoder is given by 
\begin{align} \label{eq7}
&D^{(1)}=F\left(z*X^{(3)}+B^{(3)}\right), \nonumber\\
&D^{(2)}=F\left({\rm ConvT}\left(D^{(1)},W^{(4)},B^{(4)}\right)\right), \nonumber\\
&D^{(3)}={\rm ConvT}\left(D^{(2)},W^{(5)},B^{(5)}\right).
\end{align}
Here, $D^{(3)}$ is the output of the generator, denoted by $\widehat{x}$. Let $\mathbf{W}^{\rm vae}, \mathbf{B}^{\rm vae}$ be the weight and bias matrices of the used VAE. In light of the objective (\emph{b}), the decoder reconstruction principle is cast to a minimization problem as
\begin{align*}
&\min \; \mathbb{E}\left[\left\|x-\widehat{x}\right\|^2\right], \\
&\; \emph{s.t.}\;\; \mathbf{W}^{\rm d},\; \mathbf{B}^{\rm d},
\end{align*}
where $\mathbb{E}\left[\cdot\right]$ is the expectation with respect to the variational distribution.
\smallskip

\textbf{\emph{Evidence lower-bound as loss function}.} 
Recalling the three objectives of the established VAE, one may ideally train the network with a maximum likelihood objective such that the probability assigned by the model $p_{\theta}(x_0)$ to each training example $x_0$ is as large as possible. One may thus define the following loss function:
\begin{equation*}
\mathcal{L}=D_{KL} \Big(q(z|\widehat{x})\left\|p(z)\right.\Big)+\mathbb{E}\left[\left\|x-\widehat{x}\right\|^2\right].
\end{equation*}
Yet, $p_{\theta}(x_0)$ is intractable since one has to marginalize over all the possible reverse trajectories to compute it. An alternative is to minimize a variational lower-bound of the negative log-likelihood \cite{ref-tpamidiffu}. It is the known evidence lower bound (ELBO), which is defined by 
\begin{align} \label{eq8}
\mathcal{L}^{\rm ELBO}  = & \mathbb{E}\big[-\log q_\theta(z|x) + \log p_\theta(x,z)\big]  \nonumber \\
 = & - D_{KL}\big(p(z|x)\|p(z)\big) + \mathbb{E}\big[\log p_\theta(z|x)\big]. 
\end{align}
In the following, $\mathcal{L}^{\rm ELBO}$ is used as the loss function for obtaining optimal parameters $\mathbf{W}^{\rm vae}=\{\mathbf{W}^{\rm e},\mathbf{W}^{\rm d}\}$ and $\mathbf{B}^{\rm vae}=\{\mathbf{B}^{\rm e},\mathbf{B}^{\rm d}\}$.
\smallskip

\begin{figure} 
	\centering
	\includegraphics[width=0.48\textwidth]{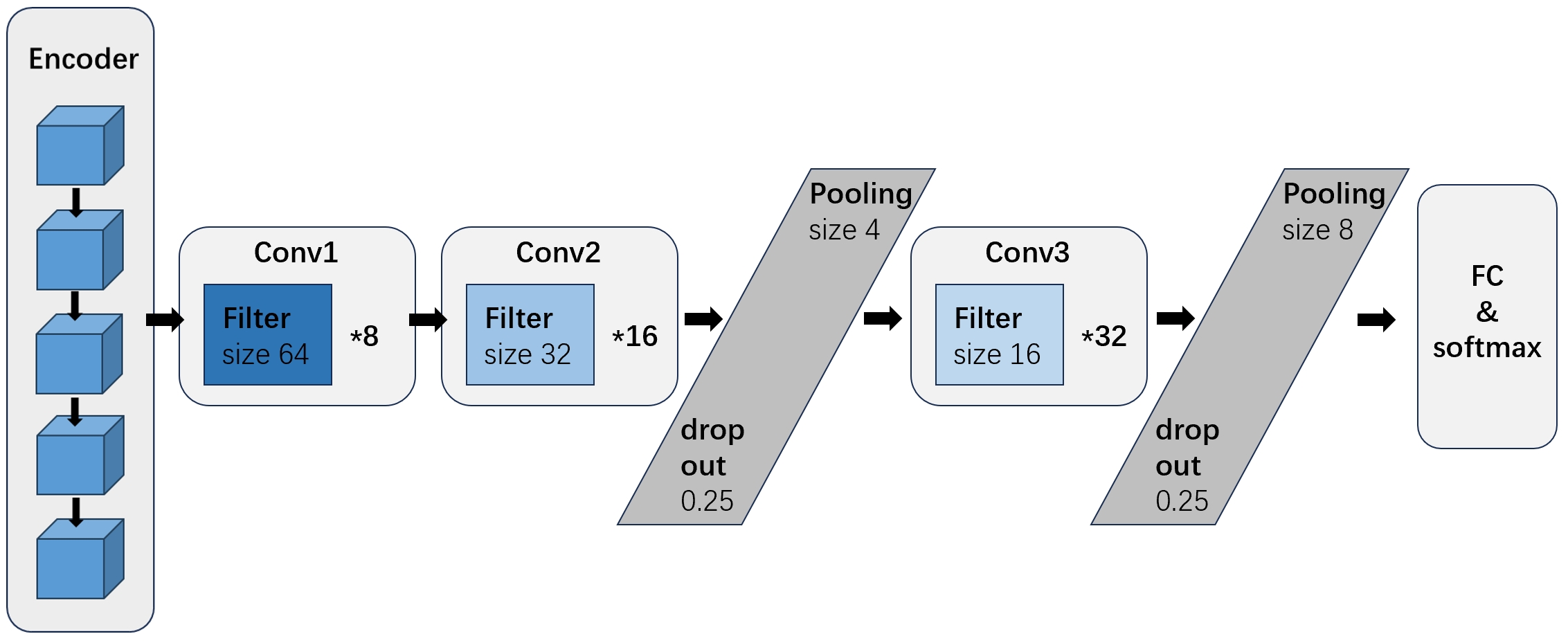}
	\caption{Illustration of data flow and sizes of the classifier. The classifier is fed with the extracted features from the encoder. Next to the 64*8 and 32*16 convolution layers, a pooling layer is set and several parameters are discarded to prevent overfitting. After that, the data are normalized using a 16*32 convolution layer and a pooling layer, and a fully connected layer plus softmax are included to yield the classification result.}\label{fig3}
\end{figure}

\textbf{\emph{Classifier for identifying depression}.} 
Next to the VAE, a classifier network (Fig. \ref{fig3}) is constructed for distinguishing depression or health. The proposed classifier uses 8 filters of size 64 on the first convolutional layer and 16 filters of size 32 on the second convolutional layer. Then, an average pooling layer of size 4 and a drop-out layer with a drop rate of 0.25 are added, so as to reduce the number of parameters and prevent the model from overfitting. The final convolutional layer consists of 32 filters of size 16, followed by an average pooling layer of size 8 and a drop-out layer with a drop rate of 0.25. Batch normalization is performed after each convolution, and an activation layer is added. This enables the network to learn and converge more quickly through regularization. The last layer outputs the classification results.

\section{Experiments}
\label{experiments}
In this section, the BIG model is evaluated on two public EEG datasets: Multimodal Open Data for Mental Disorder Analysis (MODMA) \cite{ref-modma} and STJU Emotion EEG Dataset (SEED) \cite{ref-eeg2emotion}. MODMA includes EEG signals recorded using a 128-channel HydroCel geodesic sensor network with a reference electrode at Cz, the central midline position, under a binary labeling: depression or health. SEED contains EEG signals recorded at a 1,000 Hz sampling rate with 62 channels using the ESI NeuroScan system, by a three-class labeling: negative, neutral, and positive emotion. Four types of task performances are evaluated for the BIG model: 
\begin{itemize}
	\item Completing the classification task from EEG data with comparative performances; 
    \item Assessing the computational cost advantage in reasoning; 
	\item Generating new EEGs to complement the training dataset for few-shot learning;
	\item Interpreting latent cognitive states from the generative brain network. 
\end{itemize} 

\subsection{Classification performance evaluation}\label{subsec4.1}

To assess the classification performance, EEG samples from MODMA are used as training and testing datasets. The experimental results show that the BIG model outperforms its VAE counterpart and other convolution types like EEGNet in EEG classification. Comparative works also demonstrate its computational energy-efficient advantage over other approaches like VAE and convolutional EEGNet. A similar classification task is conducted on SEED to verify its generalization. 
\smallskip

\begin{table}  [ht] 
\renewcommand\arraystretch{1.3}
\centering
\caption{\textbf{Performance comparison for identifying depression EEGs on MODMA}}\label{tab1}%
\begin{tabular}{@{\extracolsep\fill}l|ccc}
\hline
Model & Accuracy  & Specificity & Sensitivity \\
\hline
EEGNet\cite{ref-eegnet}   & 0.699   & 0.648  & 0.709  \\
TS-SEFFNet\cite{ref-tsseff}    & 0.747   & 0.661  & 0.799  \\
Deep ConvNet\cite{ref-deepconv}   &  0.762   & 0.594  &  0.913  \\
HEMAsNet\cite{ref-hemas} &  0.807   & 0.728  &  0.841  \\
BIG (Ours)  &  0.899   & 0.903  &  0.884  \\
\hline
P-value &   \multicolumn{3}{c}{$<1e-2$}   \\ 
\hline
\end{tabular}
\end{table}

\begin{table} [ht]
	\renewcommand\arraystretch{1.3}
	\centering
	\caption{\textbf{Performance comparison for identifying depression EEGs on SEED}}
	\label{tab2}
	\begin{tabular}{l|c|c}
		\hline
		Model & Accuracy  & Standard Deviation($\%$) \\
		\hline
		BiDANN-S\cite{ref-r2gstnn}   & 0.821   & 6.87  \\
		R2G-STNN\cite{ref-r2gstnn}    & 0.831   & 7.63  \\
		IAG\cite{ref-iag}   &  0.863   & 6.91  \\
		BSNN\cite{ref-bsnn} &  0.864   & 5.26  \\
		BIG (Ours)  &  0.836   & 5.44  \\
		\hline
		P-value &   \multicolumn{2}{c}{$<1e-2$}   \\ 
		\hline
	\end{tabular}
\end{table}

\begin{figure*}
	\centering
	\begin{minipage}{0.36\textwidth}
		\centering
		\includegraphics[width=\textwidth,height=0.2\textheight]{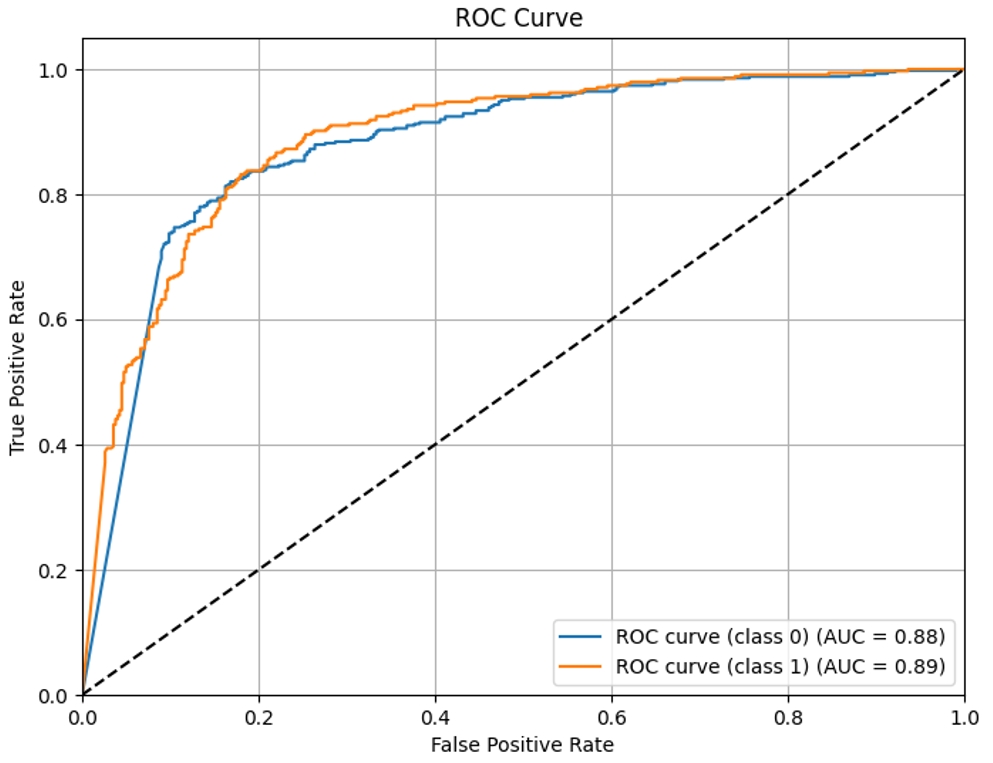} 
		\textbf{(a)} 
	\end{minipage}
	\hspace{0.4in}
	\begin{minipage}{0.36\textwidth}
		\centering
		\includegraphics[width=\textwidth,height=0.2\textheight]{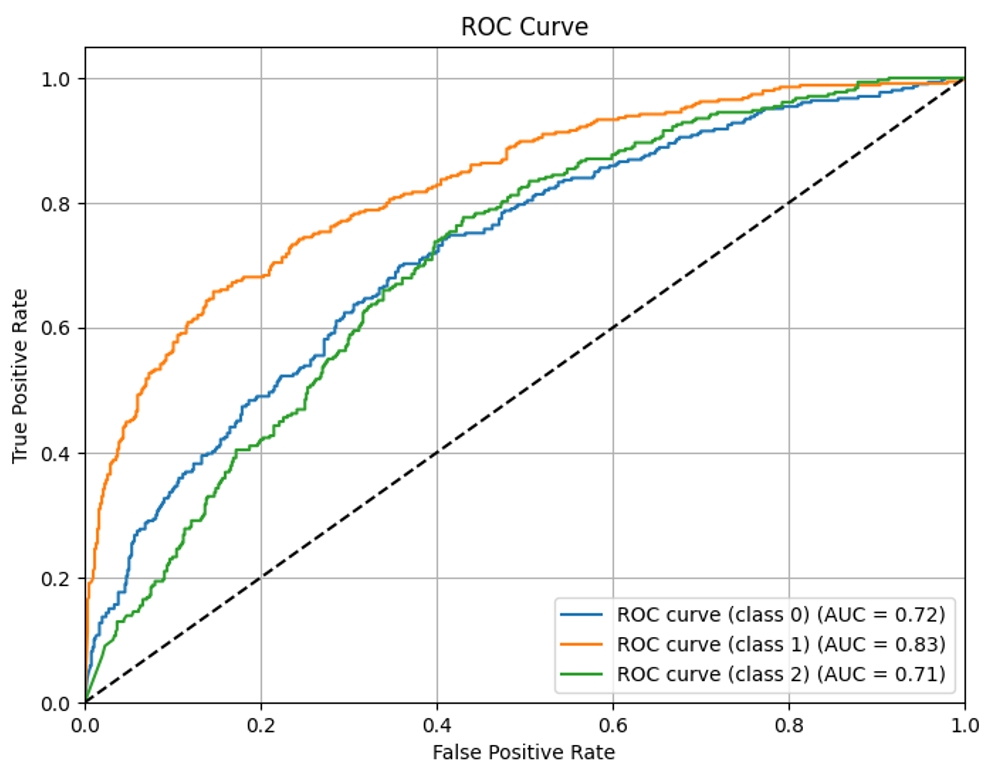} 
		\textbf{(b)} 
	\end{minipage}
	\caption{ROC curves of the BIG model on the testing datasets: (a) MODMA (two-class), (b) SEED (three-class). The model has an AUC larger than 0.88 for the two-class task, and has an AUC larger than 0.71 for the three-class task, demonstrating its sample-level robustness and generalization.}
	\label{fig4}
\end{figure*}

Table \ref{tab1} shows three metrics for evaluating the classification performance of the BIG model as compared to other networks, on the same testing dataset from MODMA. For a comprehensive evaluation, the accuracy (ACC), precision (PRE), recall (REC), and F1 score are used as metrics. The conventional VAE and Transformer are used as comparative classification models. The BIG model yields a classification accuracy of $89.9\%$, a sensitivity of $88.9\%$ and a specificity of $90.2\%$, better than the conventional networks. As can be seen in Table \ref{tab1}, the BIG model outperforms SOTAs on MODMA when used for identifying EEG signals. The BIG model yields metric improvements of 18.2\%, 0.33\%, 4.88\%, and 0.74\% respectively, compared to SOTA methods. Fig. \ref{fig4}a depicts the sample-level ROC (receiver operating characteristic) curves for the BIG model on MODMA, offering an AUC (area under curve) larger than 0.88. 
\smallskip

Furthermore, the BIG model is trained and tested with EEG samples from the SEED dataset. Table \ref{tab2} presents the classification performances and related comparisons on SEED. The BIG model achieves a classification accuracy of 83.6\%, a standard deviation of 5.44\%, exceeding some SOTA methods. Likewise, Fig. \ref{fig4}b gives the sample-level ROC of the BIG model, with an AUC bigger than 0.71. Recalling the classification performances from Table \ref{tab1} and Fig. \ref{fig4}a, enhancements in Table \ref{tab2} showcase the generalization of the BIG model when used for EEG classification. 

\subsection{Few-shot learning enhancement via generated EEGs}\label{subsec4.3}

As an enhancement of the above classification task, the BIG model can also generate EEG data since it simulates the EEG waveform trends for individual input segments. The established EEG generator is utilized to synthesize new samples to complement the original training dataset, as shown in Fig. \ref{fig2}b. 
\smallskip

\begin{figure}[bp]
	\centering
	\begin{minipage}{0.24\textwidth}
		\centering
		\includegraphics[width=\textwidth,height=0.11\textheight]{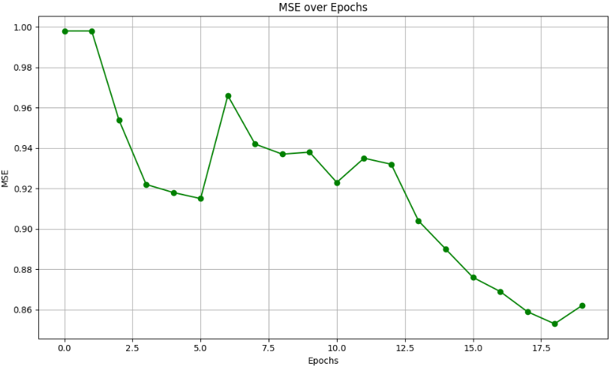} 
        \textbf{(a)}
	\end{minipage}
	\begin{minipage}{0.24\textwidth}
		\centering
		\includegraphics[width=\textwidth,height=0.11\textheight]{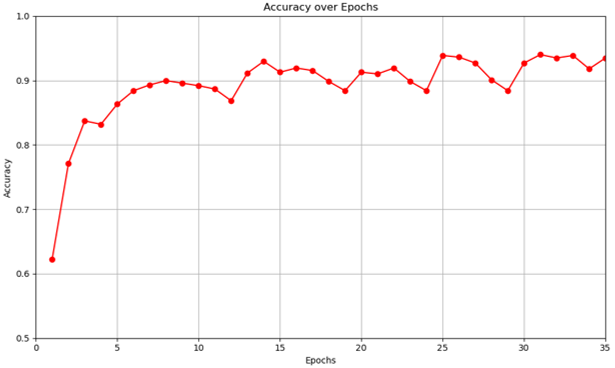} 
		\textbf{(b)} 
	\end{minipage}
	\caption{Curves of reconstruction MSE and synthesized classification accuracy. (a) {MSE of one generated EEG from BIG and one original decreases and reaches 0.86 after epoches. (b) The classification accuracy of BIG is above $90\%$ under partial training with only $20\%$ of the original samples}.}
	\label{fig6}
\end{figure}

To qualify the generative property, Fig. \ref{fig6}a depicts the MSE (mean-squared error) of the reconstruction error between the VAE-decoder outputs $\widehat{X}$ and the initial EEG samples $X$. The conventional VAE is used for comparison, to demonstrate that the BIG model has a lower error in EEG reconstruction. The newly generated EEG samples are used to extend the training dataset used in Subsection \ref{subsec4.1}. The fully and partially trained models are respectively tested for classifying the EEG data. Here, partially training means that the extended training dataset is composed of both original EEGs and newly generated EEGs. Fig. \ref{fig6}b demonstrates the classification accuracy of the BIG model, where comparisons are carried out in various scenarios of fully or partially training. Notably, {the BIG model has achieved the best accuracy of 91.86\% using only 20\% of the original samples}. 
The MSE and accuracy curves indicate that the reconstructed EEG samples are usable and reliable, supporting one generative AI solution to few-shot learning with the BIG model. 

\subsection{Network dynamic sources of cognitive states}\label{subsec4.4}

To identify potential sources of cognitive states, experiments are conducted to assess whether the BIG model reflects brain-inspired neuronal population activations beneath the EEG data. Specifically, the established generative brain network is evaluated using the impulse activation and the functional connectivity on the MODMA dataset. 
\smallskip

\begin{figure}[htbp]
	\centering
	\begin{minipage}{0.4\textwidth}
		\centering
		\includegraphics[width=\textwidth]{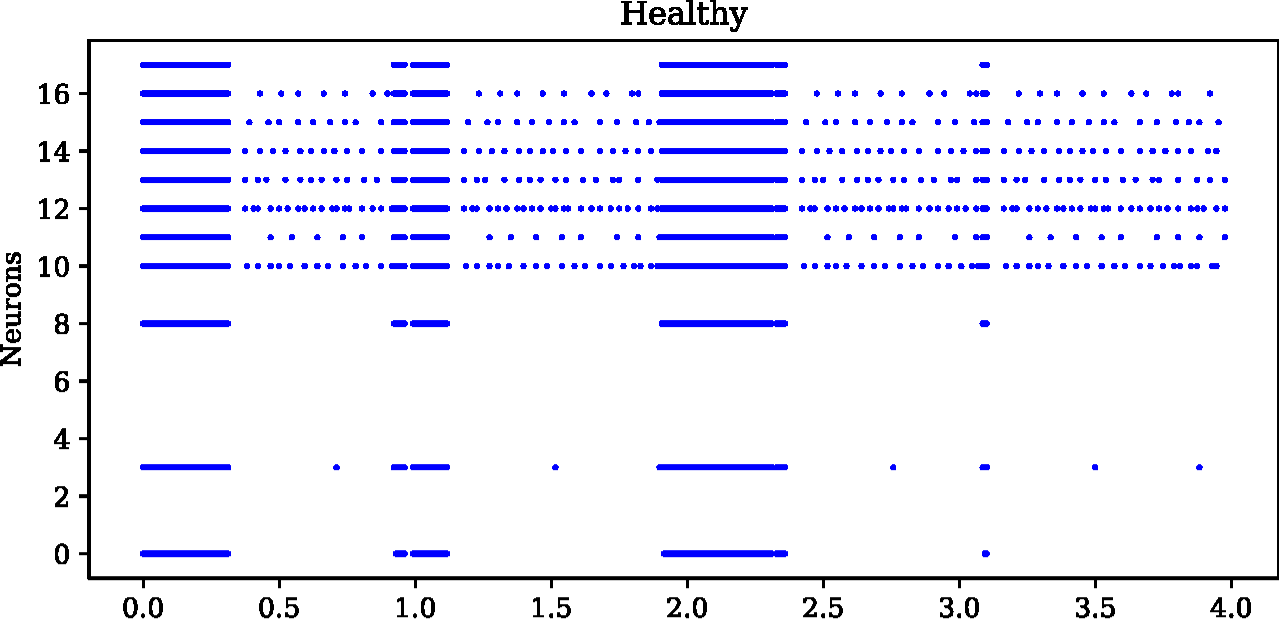} 
		\textbf{(a)} 
	\end{minipage}
	\begin{minipage}{0.4\textwidth}
		\centering
		\includegraphics[width=\textwidth]{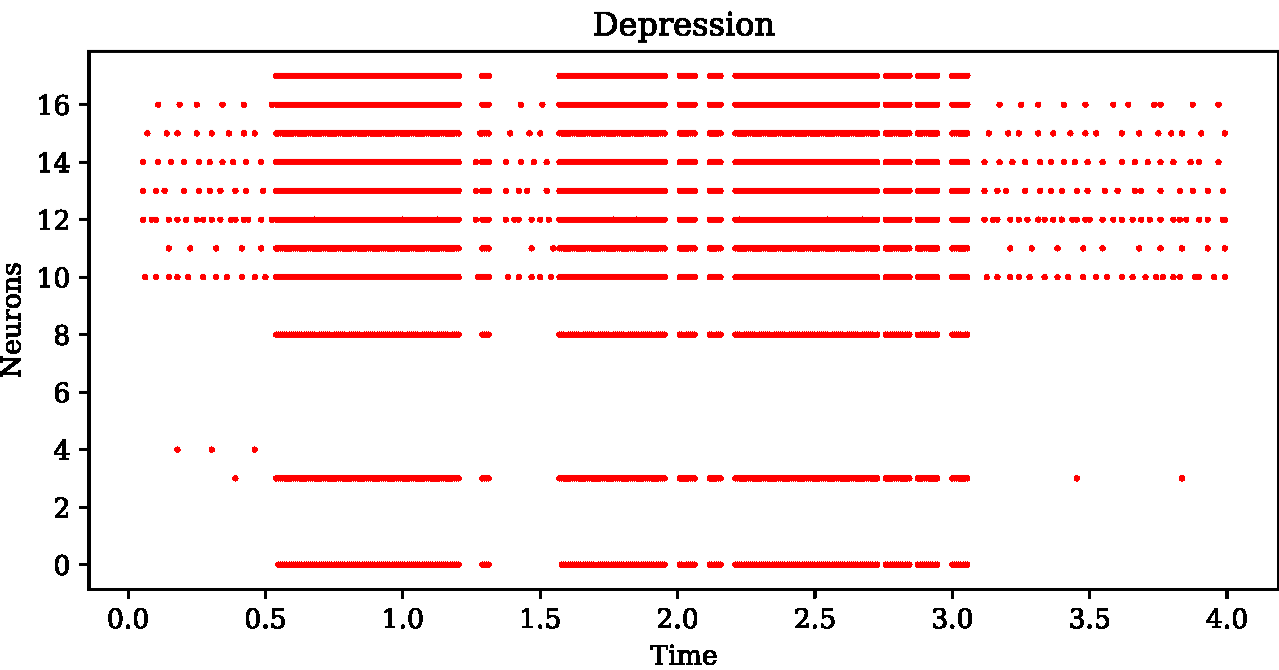} 
		\textbf{(b)} 
	\end{minipage}
	\begin{minipage}{0.4\textwidth}
		\centering
		\includegraphics[width=\textwidth]{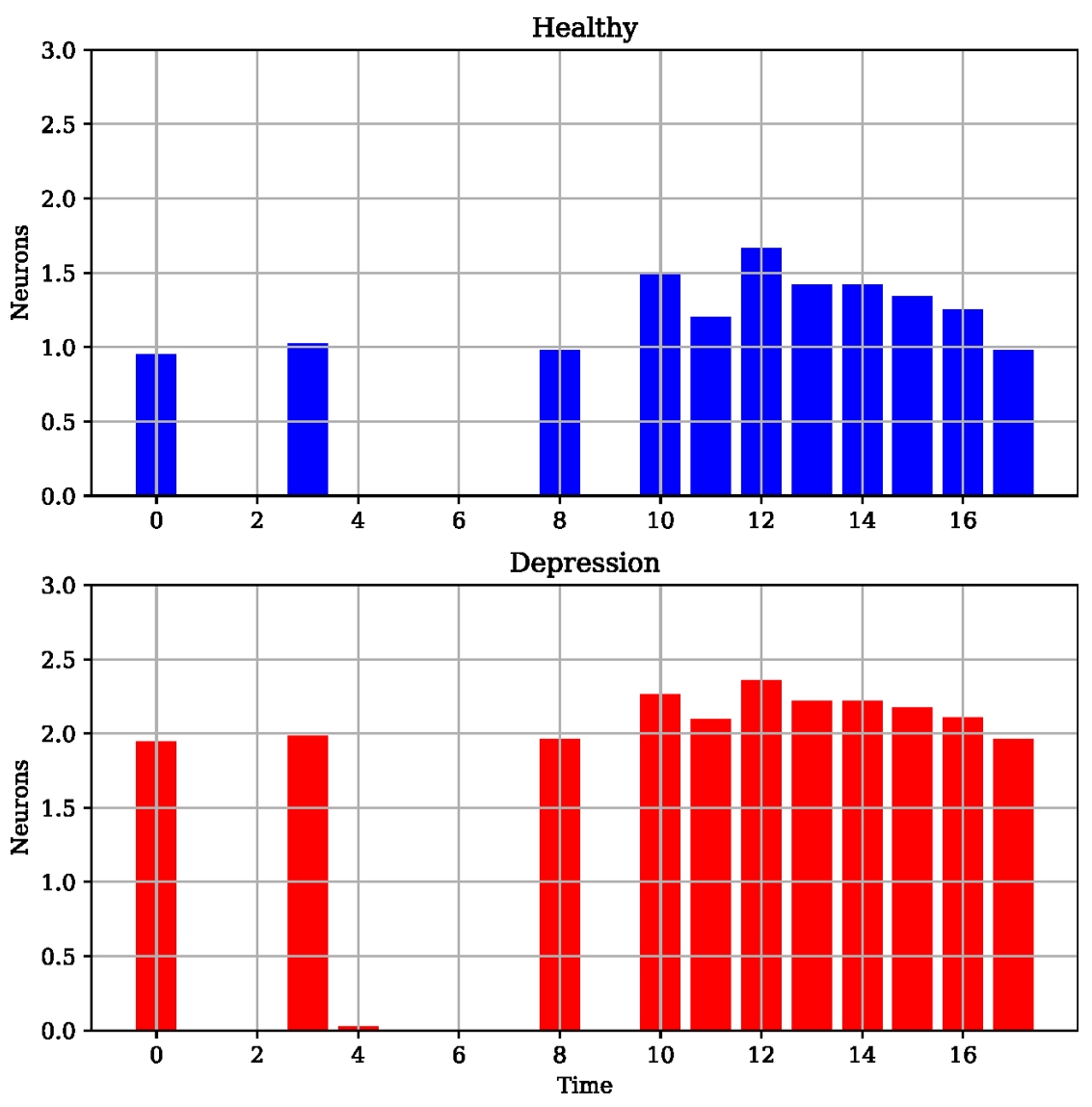} 
		\textbf{(c)} 
	\end{minipage}
	\caption{Firing rate distribution of 18 randomly chosen neurons under the impulse estimate. Panels (a) and (b) present the raster plots for healthy and depressive individuals, respectively. The firing rate of the depressive neurons is clearly higher than that of the healthy ones, and the depressive ones have a longer refractory period after firing. Panel (c) shows the histograms in terms of the channel-wise firing rate for healthy neurons (blue) and depressive ones (red).}
	\label{fig7}
\end{figure}

\textbf{\emph{Locating neuronal activations over the brain}.} 
Fig. \ref{fig7}a shows a raster plot of the impulse activity from 18 randomly chosen neurons with respect to a depressive EEG. Fig. \ref{fig7}b presents a counterpart raster plot from a healthy EEG. Combining the two raster plots, it can be seen that depressive neurons are firing more regularly when channel-norm is applied. Fig. \ref{fig7}c also indicates a histogram comparison of the firing rates in each channel respectively from depressive and healthy EEGs. It is evident that depressive EEG produces a much higher firing rate in majority of the channels. Particularly in channel 2, channel-norm produces a firing rate 20 times higher than that of the layer-norm. The three plots in Fig. \ref{fig7} suggest that the impulse estimate is reliable for precisely locating the channel-wise abnormal neuronal activations. 
\smallskip

\begin{figure*}[htbp]
	\centering
	\begin{minipage}{0.22\textwidth}
		\centering
		\includegraphics[width=\textwidth]{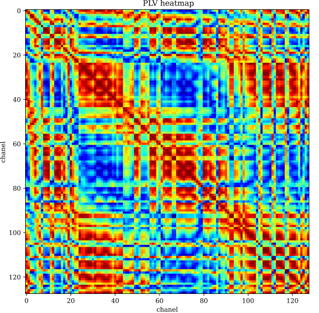} 
		\textbf{(a)} 
	\end{minipage}
	\hfill
	\begin{minipage}{0.22\textwidth}
		\centering
		\includegraphics[width=\textwidth]{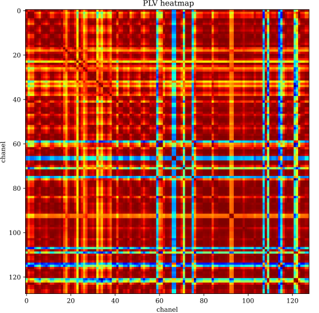} 
		\textbf{(b)} 
	\end{minipage}
	\hfill
	\begin{minipage}{0.22\textwidth}
		\centering
		\includegraphics[width=\textwidth]{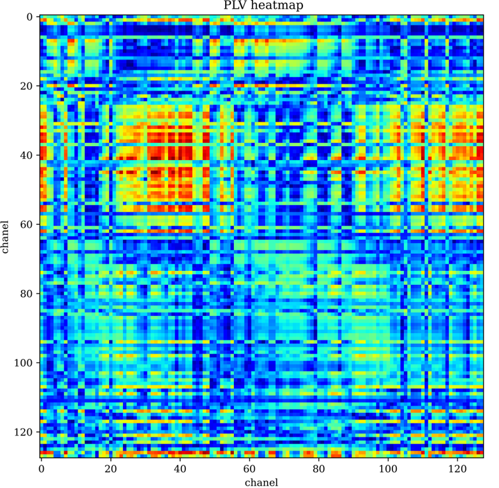} 
		\textbf{(c)} 
	\end{minipage}
	\hfill
    \begin{minipage}{0.25\textwidth}
	    \centering
    	\includegraphics[width=\textwidth]{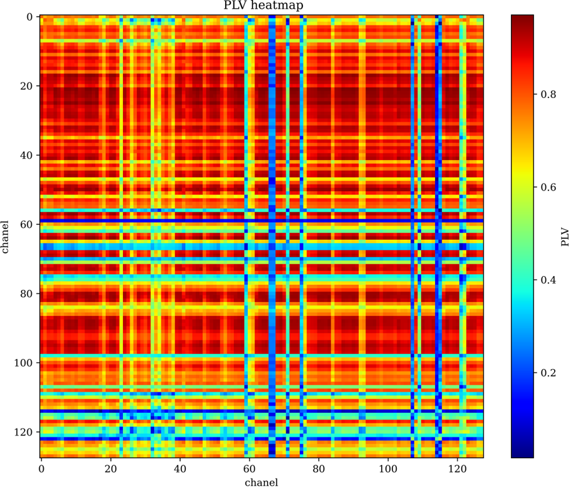} 
    	\textbf{(d)} 
    \end{minipage}
	\caption{Visualization of PLV between the original and reconstructed EEG samples of depressive and healthy types. (a) Generated healthy EEG samples. (b) Generated depressed EEG samples. (c) Original healthy EEG samples. (d) Original depressed EEG samples. The data generated by the BIG model has a high similarity with the original samples in terms of key node correlation, and the rest of the data maintains diversity to ensure that the training results are generalizable.}
	\label{fig8}
\end{figure*}

\begin{figure*}[htbp]
	\centering
	\begin{minipage}{0.18\textwidth}
		\centering
		\includegraphics[width=\textwidth]{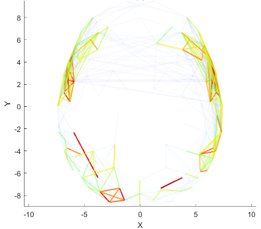} 
		\textbf{(a)} 
	\end{minipage}
	\hfill
	\begin{minipage}{0.18\textwidth}
		\centering
		\includegraphics[width=\textwidth]{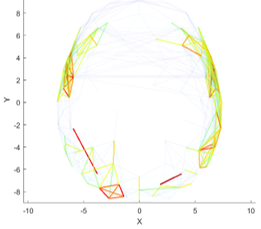} 
		\textbf{(b)} 
	\end{minipage}
	\hfill
	\begin{minipage}{0.18\textwidth}
		\centering
		\includegraphics[width=\textwidth]{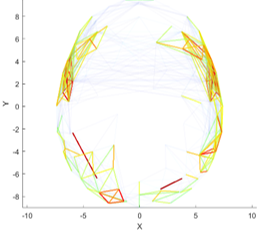} 
		\textbf{(c)} 
	\end{minipage}
	\hfill
	\begin{minipage}{0.18\textwidth}
		\centering
		\includegraphics[width=\textwidth]{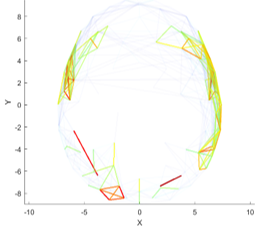} 
		\textbf{(d)} 
	\end{minipage}
	\hfill
    \begin{minipage}{0.18\textwidth}
	    \centering
	    \includegraphics[width=\textwidth]{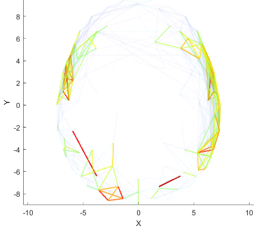} 
	    \textbf{(e)} 
    \end{minipage}   
	\begin{minipage}{0.18\textwidth}
	\centering
	\includegraphics[width=\textwidth]{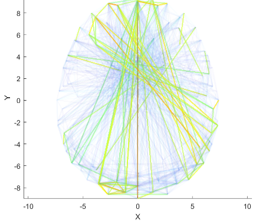} 
	\textbf{(f)} 
    \end{minipage}
    \hfill
    \begin{minipage}{0.18\textwidth}
	    \centering
	    \includegraphics[width=\textwidth]{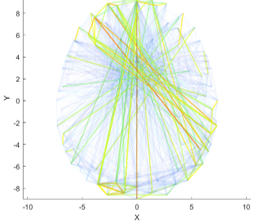} 
    	\textbf{(g)} 
    \end{minipage}
    \hfill
    \begin{minipage}{0.18\textwidth}
	    \centering
    	\includegraphics[width=\textwidth]{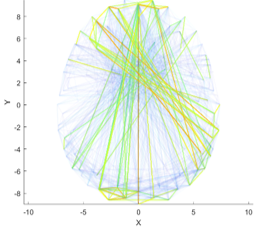} 
    	\textbf{(h)} 
    \end{minipage}
    \hfill
    \begin{minipage}{0.18\textwidth}
    	\centering
    	\includegraphics[width=\textwidth]{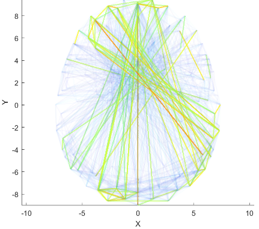} 
    	\textbf{(i)} 
    \end{minipage}
    \hfill
    \begin{minipage}{0.18\textwidth}
    	\centering
    	\includegraphics[width=\textwidth]{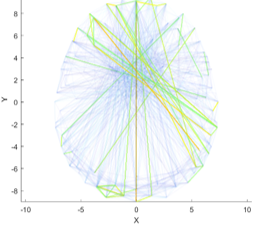}
    	\textbf{(j)} 
    \end{minipage}
	\caption{PLV projection onto the cortical space of the brain. Panels (a-e) show the dynamic changes in node correlation during the original EEG process (1min slice, 5min in total), and Panels (f-g) show the changes in the generated EEG process. The BIG model and the original data both have dynamic activation performances in the right occipital lobe and left frontal lobe throughout the process.}
	\label{fig9}
\end{figure*}

\textbf{\emph{Predicting functional connectivity.}} 
Fig. \ref{fig8} demonstrates the PLV comparison between the original and reconstructed EEG samples of healthy and depressive types, using the calculation. Panels (a) and (b) respectively illustrate the full PLV matrices corresponding to reconstructed EEG, while panels (c) and (d) present the counterparts for the original EEG. From the four plots, it can be observed that the PLV matrix generated with the original EEG signal is highly similar to the counterpart in key node connectivity. Furthermore, the PLV analysis results are projected back onto a cortical surface, from which the scores can be used to sort the input connection groups to generate a whole-brain connection map. Fig. \ref{fig9} provides an illustration.  
\smallskip

While being subject to weak negative signal disturbances, neurons that remain silent under normal thresholds will be activated. This phenomenon might lead to abnormal impulse bursting in certain brain regions, causing depressive symptoms \cite{ref-brainnetanomaly,ref-biodisorder}. Likewise, the found depressive brain regions (Fig. \ref{fig9}) reconcile to those recognized ones, e.g., prefrontal cortex dysfunction, right frontal lobe atrophy, and left inferior parietal lobe activity dysregulation in depressive patients \cite{ref-brainmdd}. Theoretically, a higher degree of firing rate indicates an increase trend of the functional connectivity (Fig. \ref{fig7} and Fig. \ref{fig8}), reconciling to brain functional networks reported in the neuroscience literature \cite{ref-plvnormal}. Overall, the BIG model holds computational interpretability that shows overlap with brain network theories and clinical findings.

\section{Conclusion}
\label{conclusion}
In this article, a brain-inspired generative (BIG) model has been developed for recognizing and analyzing EEG data, serving for the purposes of classification, generation and interpretation. The feedforward network consists of three modules: an impulsive-attention neural network, a VAE, and a generative brain network. The training process utilizes a hybrid method of gradient-based backpropagation learning and heteroassociative memory. The BIG model shows a balance of comparable EEG classification accuracy with prominently lower computational cost. As a generative model, the BIG model is capable of simulating and reproducing EEG signals, which complements the training samples for few-shot learning. By estimating the firing rate of neurons in IANN and the PLV matrix, the generative brain network further provides interpretable evidences for EEG-based depression diagnosis --- identifying abnormalities in both neuronal impulse activations and functional connectivity. 
\smallskip

\section*{Acknowledgments}
The authors gratefully acknowledge the review \& algorithm help provided by Bokun Zhang and Longhui Zhou, and the review \& editing help provided by Prof. Guanrong Chen.

\vfill


\begin{thebibliography}{1}
\bibliographystyle{IEEEtran}

\bibitem{ref-digtbrain} 
K. Amunts, M. Axer, S. Banerjee, {\em et al.}, 
The coming decade of digital brain research: A vision for neuroscience at the intersection of technology and computing. \emph{Imaging Neuroscience}, 2024, 2: 1--35.

\bibitem{ref-bscale}
W. Lu, L. Zeng, J. Wang, S. Xiang, Y. Qi, Q. Zheng, N. Xu, and J. Feng,
Imitating and exploring the human brain's resting and task-performing states via brain computing: scaling and architecture. \emph{National Science Review}, 2024, 11(5): nwae080.

\bibitem{ref-tpamidiffu}
F.-A. Croitoru, V. Hondru, R. T. Ionescu, and M. Shah, 
Diffusion models in vision: A survey. \emph{IEEE Transactions on Pattern Analysis and Machine Intelligence}, 2023, 45(9): 10850--1869. 

\bibitem{ref-gmai} 
M. Moor, O. Banerjee, Z. S. H. Abad, {\em et al.}, 
Foundation models for generalist medical artificial intelligence. \emph{Nature}, 2023, 616: 259--265.

\bibitem{ref-shireason}
G. Yang, S. Wang, J. Yang, and P. Shi, 
Desire-driven reasoning considering personalized care preferences.
\emph{IEEE Transactions on Systems, Man, and Cybernetics: Systems}, 2022, 52(9): 5758--5769.

\bibitem{ref-wanghealth}
W. Yue, Z. Wang, J. Zhang, and X. Liu,
An overview of recommendation techniques and their applications in healthcare.
\emph{IEEE/CAA Journal of Automatica Sinica}, 2021, 8(4): 701--717.

\bibitem{ref-mlreview}
M. M. Ahsan, S. A. Luna, and Z. Siddique,  
Machine-learning-based disease diagnosis: A comprehensive review. \emph{Healthcare}, 2022, 10(3): 541.

\bibitem{ref-spikereview}
J. Jordan, M. Schmidt, W. Senn, and M. A. Petrovici, 
Evolving interpretable plasticity for spiking networks. \emph{Elife}, 2021, 10: e66273.

\bibitem{ref-eeg2speech}
J. Zhou, Y. Duan, Y. Zou, Y.-C. Chang, Y.-K. Wang, and C.-T. Lin,
Speech2EEG: Leveraging pretrained speech model for EEG signal recognition. \emph{IEEE Transactions on Neural Systems and Rehabilitation Engineering}, 2023, 31: 2140--2153.

\bibitem{ref-modma}
H. Cai, Z. Yuan, Y. Gao, \emph{et al.}, 
A multi-modal open dataset for mental-disorder analysis. \emph{Scientifc Data}, 2022, 9, 178.  

\bibitem{ref-eeg2emotion}
Y. X. Wang, S. Qiu, D. Li, C. D. Du, B.-L. Lv, and H. G. He, 
Multi-modal domain adaptation variational autoencoder for EEG-based emotion recognition. \emph{IEEE/CAA Journal of Automatica Sinica}, 2022, 9(9): 1612--1626.

\bibitem{ref-linkjzhang}
K. Zhang, J. Shen, G. He, Y. Sun, H. Ling, H. Zha, H. Li, and J. Zhang, 
A transformative topological representation for link modeling, prediction and cross-domain network analysis. \emph{IEEE Transactions on Pattern Analysis and Machine Intelligence}, 2024, 46(9): 6126--6138. 

\bibitem{ref-biodisorder}
T. Prevot and E. Sibille,  
Altered GABA-mediated information processing and cognitive dysfunctions in depression and other brain disorders. \emph{Molecular Psychiatry}, 2021, 26(1): 151-167.

\bibitem{ref-cortical1}
S. H. Journee, V. P. Mathis, C. Fillinger, P. Veinante, and I. Yalcin,
Janus effect of the anterior cingulate cortex: Pain and emotion. \emph{Neuroscience and Biobehavioral Reviews}, 2023, 153: 105362.

\bibitem{ref-brainmdd}
B. D. Hare and R. S. Duman,  
Prefrontal cortex circuits in depression and anxiety: contribution of discrete neuronal populations and target regions. \emph{Molecular Psychiatry}, 2020, 25(11): 2742--2758.

\bibitem{ref-cortical2}
M. Ullsperger and O. Stork,
To err is (not only) human: Mechanisms of post-error attentional regulation illuminated in mice. \emph{Neuron}, 109: 1074--1076.

\bibitem{ref-networkurths}
Y. Zou, R. V. Donner, N. Marwan, J. F. Donges, and J. Kurths, 
Complex network approaches to nonlinear time series analysis. \emph{Physics Reports}, 2019, 787: 1--97.

\bibitem{ref-spikeatt}
Q. Zhou, Z. Huang, M. Ding, and X. Zhang,
Medical image classification using light-weight CNN with spiking cortical model based attention module. \emph{IEEE Journal of Biomedical and Health Informatics}, 2023, 27(4): 1991-2002.

\bibitem{ref-brainnetanomaly}
A. Bessadok, M. A. Mahjoub, and I. Rekik, 
Graph neural networks in network neuroscience. \emph{IEEE Transactions on Pattern Analysis and Machine Intelligence}, 2023, 45(9): 10850--1869. 

\bibitem{ref-ourmemory}
B.~Hu, Z.-H. Guan, G.~Chen, and F.~L. Lewis, 
Multistability of delayed hybrid impulsive neural networks with application to associative memories. \emph{IEEE Transactions on Neural Networks and Learning Systems}, 2019, 30(5): 1537--1551. 

\bibitem{ref-memory}
B. Kosko, 
Bidirectional associative memories: Unsupervised Hebbian learning to bidirectional backpropagation. \emph{IEEE Transactions on Systems, Man, and Cybernetics: Systems}, 2021, 51(1): 103--115. 

\bibitem{ref-ganbrainima} 
X. Gao, F. Shi, D. Shen, and M. Liu, 
Task-induced pyramid and attention GAN for multimodal brain image imputation and classification in Alzheimer's disease. \emph{IEEE Journal of Biomedical and Health Informatics}, 26, 36--43, 2021.

\bibitem{ref-vaeimage}
A. Alfakih, Z. Xia, B. Ali, S. Mamoon, and J. Lu,
Deep causality variational autoencoder network for identifying the potential biomarkers of brain disorders. \emph{IEEE Transactions on Neural Systems and Rehabilitation Engineering}, 32: 112--121, 2023.

\bibitem{ref-eeg2vec}
D. Bethge, P. Hallgarten, T. Grosse-Puppendahl, \emph{et al.}, 
EEG2Vec: Learning affective EEG representations via variational autoencoders. \emph{2022 IEEE International Conference on Systems, Man, and Cybernetics}, 2022, pp: 3150--3157.

\bibitem{ref-graphvae}
T. Behrouzi and D. Hatzinakos,
Graph variational auto-encoder for deriving EEG-based graph embedding. \emph{Pattern Recognition}, 2022, 121: 108202.

\bibitem{ref-cogvae}
D. D. Chakladar, S. Datta, P. P. Roy, and V. A. Prasad,
Cognitive workload estimation using variational autoencoder and attention-based deep model. \emph{IEEE Transactions on Cognitive and Developmental Systems}, 15(2): 581--590, 2023. 

\bibitem{ref-eegnet}
V. J. Lawhern, A. J. Solon, N. R. Waytowich, S. M. Gordon, C. P. Hung, and B. J. Lance, 
{EEGN}et: a compact convolutional neural network for {EEG}-based brain--computer interfaces. \emph{Journal of Neural Engineering}, 2018, 15(5): 056013. 

\bibitem{ref-tsseff}
Y. Li, L. Guo, Y. Liu, J. Liu, and F. Meng, 
A temporal-spectral-based squeeze-and-excitation feature fusion network for motor imagery EEG decoding. \emph{IEEE Transactions on Neural Systems and Rehabilitation Engineering}, 2021, 29(8): 1534–1545.

\bibitem{ref-deepconv}
R. T. Schirrmeister, J. T. Springenberg, L. D. J. Fiederer, \emph{et al.}, 
Deep learning with convolutional neural networks for EEG decoding and visualization. \emph{Human Brain Mapping}, 2017, 38: 5391--5420.

\bibitem{ref-hemas}
J. Shen, K. Li, H. Liang, Z. Zhao, Y. Ma, J. Wu, J. Zhang, Y. Zhang, and B. Hu, 
HEMAsNet: A hemisphere asymmetry network inspired by the brain for depression recognition from electroencephalogram signals. \emph{IEEE Journal of Biomedical and Health Informatics}, 2024, 28(9): 5247--5259.

\bibitem{ref-eegcnnmdd}
A. Seal, R. Bajpai, J. Agnihotri, \emph{et al.}, 
DeprNet: A deep convolution neural network framework for detecting depression using EEG. \emph{IEEE Transactions on Instrumentation and Measurement}, 2021, 70: 1-13.

\bibitem{ref-treemdd}
D. Colledani, P. Anselmi, and E. Robusto, 
Machine learning-decision tree classifiers in psychiatric assessment: An application to the diagnosis of major depressive disorder. \emph{Psychiatry Research}, 2023, 322: 115127. 

\bibitem{ref-learnepi}
B. Abbasi and D. M. Goldenholz,  
Machine learning applications in epilepsy. \emph{Epilepsia}, 2019, 60(10): 2037-2047.

\bibitem{ref-hybridepi}
A. Subasi, J. Kevric, and C. M. Abdullah,  
Epileptic seizure detection using hybrid machine learning methods. \emph{Neural Computing and Applications}, 2019, 31: 317-325.

\bibitem{ref-atteeg} 
E. Q. Wu, Y. Gao, W. Tong, Y. Hou, R. Law, and G. Zhu,
Cognitive state detection in task context based on graph attention network during flight.
\emph{IEEE Transactions on Systems, Man, and Cybernetics: Systems}, 2023, 54(9): 5224--5236. 

\bibitem{ref-huangsnn} 
Y. Zheng, L. Zheng, Z. Yu, T. Huang, and S. Wang,
Capture the moment: High-speed imaging with spiking cameras through short-term plasticity. \emph{IEEE Transactions on Pattern Analysis and Machine Intelligence}, 45(7): 8127--8142, 2023. 

\bibitem{ref-ourtai} 
B.~Hu, Z.-H. Guan, G.~Chen, and J. Kurths,
Energy-efficient hybrid impulsive model for joint classification and segmentation on CT images. \emph{IEEE Transactions on Artificial Intelligence}, DOI: 10.1109/TAI.2024. 3517570, 2024: 1--13.

\bibitem{ref-spiketran} 
M. Yao, J. Hu, Z. Zhou, L. Yuan, Y. Tian, B. Xu, and G. Li, 
Spike-driven Transformer. \emph{2023 Conference on Neural Information Processing Systems}, 2023, pp: 1--16.

\bibitem{ref-glocal}
Y. Wu, R. Zhao, J. Zhu, \emph{et al}., 
Brain-inspired global-local learning incorporated with neuromorphic computing. \emph{Nature Communications}, 13: 65, 2022. 

\bibitem{ref-loihi}
M. Davies, N. Srinivasa, T. H. Lin, \emph{et al}.,  
Loihi: A neuromorphic manycore processor with on-chip learning. \emph{IEEE Micro}, 2018, 38(1): 82-99.

\bibitem{ref-spikeeg}
Y. Yang, J. K. Eshraghian, N. D. Truong, \emph{et al}.,
 Neuromorphic deep spiking neural networks for seizure detection. \emph{Neuromorphic Computing and Engineering}, 2023, 3(1): 014010.

\bibitem{ref-fewsnn}
R. Jiang, J. Zhang, R. Yan, and H. J. Tang,  
Few-shot learning in spiking neural networks by multi-timescale optimization. \emph{Neural Computation}, 2021, 33(9): 2439-2472.

\bibitem{ref-fewspike} 
A. Bittar and P. N. Garner, 
A surrogate gradient spiking baseline for speech command recognition. \emph{Frontiers in Neuroscience}, 2022, 16: 865897.

\bibitem{ref-neurosnn}
G. Indiveri and S.-C. Liu, 
Memory and information processing in neuromorphic systems, \emph{Proceedings of the IEEE}, 2015, 103(8): 1379--1397.

\bibitem{ref-plvnormal}
S. D. Kulik, L. Douw, E. van Dellen, \emph{et al.}, 
Comparing individual and group-level simulated neurophysiological brain connectivity using the Jansen and Rit neural mass model. \emph{Network Neuroscience}, 2023, 7(3): 1--15.

\bibitem{ref-r2gstnn}
Y. Li, W. Zheng, L. Wang, Y. Zong, and Z. Cui, 
From regional to global brain: a novel hierarchical spatial-temporal neural network model for EEG emotion recognition. \emph{IEEE Transactions on Affective Computing}, 2022, 13(2): 568--578. 

\bibitem{ref-iag}
T. Song, S. Liu, W. Zheng, Y. Zong, and Z. Cui, 
Instance-adaptive graph for EEG emotion recognition. \emph{Proceedings of the AAAI Conference on Artificial Intelligence}, 2020, pp. 2701--2708.

\bibitem{ref-bsnn}
H. Chang, Y. Zong, W. Zheng, C. Tang, J. Zhu, and X. Li, 
Depression assessment method: An EEG emotion recognition framework based on spatiotemporal seural network. \emph{Frontiers in Psychiatry}, 2022, 13:  851589.

\end{thebibliography}
\end{document}